**Emergence de la complexité dans un modèle simple de comportement mécanique des roches**

**Emerging complexity in a simple model of the mechanical behavior of rocks**

David AMITRANO, LAEGO-INPL, Ecoles des Mines de Nancy

Parc de Saurupt, 54024 Nancy Cedex

amitrano@mines.inpl-nancy.fr

**Résumé :** Nous présentons un modèle simple pour le comportement mécanique des roches, basé sur un endommagement progressif et une interaction élastique entre éléments. Ce modèle permet de simuler un certain nombre d'observations expérimentales : comportement mécanique allant du fragile au ductile, structure fractale de l'endommagement, distribution en loi puissance des avalanches d'endommagement. Ces propriétés macroscopiques ne sont pas introduites à l'échelle élémentaire mais résultent de l'interaction entre éléments. L'émergence de cette complexité permet de considérer le processus de déformation des roches comme un système complexe possédant une dynamique non-linéaire.

**Abstract :** We propose a mechanical model for the behavior of rocks based on progressive damage at the elementary scale and elastic interaction. It allows us to simulate several experimental observations: mechanical behavior ranging from brittle to ductile, fractal structure of the damage, power-law distribution of the damage avalanches. These macroscopic properties are not incorporated at the elementary scale but are the results of the interaction between elements. This emerging complexity permits us to consider the strain rock process as a complex system characterized by non-linear dynamics.

**Abridged version**

# 1. Introduction

For scales ranging from laboratory samples to the Earth crust, the mechanical behavior of rocks is often considered as complex, term generally used instead of complicated. This complexity is involved by the various mechanical behaviors which are observed for rocks (elastic/plastic, brittle/ductile, damage, viscosity,... ) and to the numerous parameters influencing the behavior which are related to the rock fabric (mineralogy, microstructure, crack density and distributions,...) and/or to the loading conditions (pressure, temperature, strain rate) [13]. The complexity is also related to the low scale strain mechanisms (dislocation flow, cracks propagation, fracture shearing,...) which interact and whose the effects on the macroscopic behavior are difficult to estimate.

The complexity is also revealed by the dynamics of the deformation process, e.g. as observed through the seismic waves emitted by the low scale processes (acoustic emission at the laboratory scale or seismicity at the earth crust scale). On a wide range of scales, the seismicity displays both an unpredictability of each event (size, time and location of occurrence) and a statistical predictability. As a matter of fact, one may observe particularly robust power-law distributions on both size, time and space domains [1, 2, 8, 11, 12, 17, 18, 20-23, 25, 26] which indicate scaling invariance and long range correlation. The fractal properties observed for damage [3, 9, 15, 24] (roughness, microstructure) are supplementary arguments for complexity of the rock deformation.

In order to better understand what is the origin of this complexity, we propose a numerical model based on a simple elementary behavior which is able to reproduce numerous experimental observations at macroscopic scale. Hence the behavior of rocks can be considered as resulting from emerging complexity, i.e. the macroscopic properties are not incorporated at the elementary scale but are the results of the interaction between elements.

## 2. Model features

The proposed model is based on progressive elastic damage. The effective elastic modulus of the damaged material, $E_{eff.}$, is expressed as a function of the initial modulus, $E_0$, and the damage $D$.

$$E_{eff.} = E_0 \cdot (1-D) \qquad (1)$$

D is quantitatively related to the cracks density [14]. Such a relation works when the considered volume is large compared with the defect size (i.e. cracks size). Hence, the elementary scale is mesoscopic. The simulated material is discretized using a 2D finite element method with plane strain hypothesis. The loading consists on progressively increasing the vertical displacement of the upper boundary of the model. When an element exceeds a given damage threshold, its elastic modulus is multiplied by a factor (1-D). Because of the elastic interaction, the stress redistribution can induce an avalanche of damages. The number of damaged elements during a single loading step is the avalanche size. The Mohr-Coulomb criterion is used as a damage threshold.

$$\tau = \sigma \tan \phi + C \qquad (2)$$

Where $\tau$ is the shear stress; $\sigma$ is the normal stress; $\phi$ is the internal friction angle and C is the cohesion. In order to obtain a macroscopic behavior which differs from the elementary one, it is necessary to integrate heterogeneity in the elements properties [1, 2]. Here we present results for which the cohesion is randomly drawn from a uniform distribution. More details on the model feaures are given in references [1, 2].

## 3. Macroscopic properties

The study of the model sensitivity has shown that the macroscopic behavior is only controlled by the internal friction angle $\phi$ [1, 2]. Figure 1 displays simulation results for $\phi$ ranging from 0 to 56°, i.e. the macroscopic behavior, the final damage state, the size frequency distribution and the spatial correlation integral of the damage C(r) [6]. The correlation dimension $D_2$ is calculated by estimating the slope of C(r) in a loglog diagram. The fitting is restricted to the power-law section of C(r).

The first stage of the mechanical behavior is linear. The non linearity appears with the onset of the damage activity. Regarding the value of $\phi$, the behavior ranges continuously from ductile to brittle (figure 1a). In the brittle case, there is a huge and sudden stress decrease just after the peak which could be considered as a macrorupture of the model. This is associated with the damage localization within thin shear bands (figure 1b). On the opposite, the ductile behavior is characterized by the absence of macroscopic stress decrease. The damage becomes progressively localized within thick shear bands. All the intermediary situations can be reached by tuning the $\phi$ parameter. The size frequency distributions follows a power-law whose the exponent (slope on a loglog plot) depends on the $\phi$ parameter (figure 1c). The integral $C(r)$ follows a power law (figure 1d), what indicates a fractal structure of the damage. The correlation dimension $D_2$ depends on the $\phi$ parameter.

All the macroscopic properties are sensible to the $\phi$ parameter. The study of the stress field around a single defect has shown that this parameter strongly influences the interaction geometry between elements [1, 2]. The higher the $\phi$ parameter is, the more anisotropic the interaction is. Hence this anistropy influences both the damage localisation, the avalanches dynamics and the macroscopic behavior.

**4. Emerging complexity and implication for the dynamics of rock deformation**

The proposed model is based on an elementary progressive damage within an elastic heterogeneous model. Each element has isotropic properties associated with a simple behavior, i.e. decrease of the elastic modulus by discrete damage events. At the macroscopic scale, one may observe different aspects of a complex behavior: the mechanical behavior is non-linear and ranges from ductile to brittle, the final damage state has a fractal structure, the size-frequency of damage events follows a power-law. The figure 2 gives a synthetic scheme of all these observations. As these properties are not incorporated at the elementary scale, they are emerging properties of the system due to the interaction between elements. We have shown that changing the geometry of interaction (by changing the ϕ parameter) modifies all the macroscopic properties. According to that, the simulated deformation process can be considered as a complex system [4, 10, 16]. The emergence of free-scale distributions for both the size and space distribution is a supplementary aspect of complexity [4].

## 5. Conclusion

A simple mechanical model has been proposed for the behavior of rocks. It permits us to simulate several experimental observations: mechanical behavior ranging from brittle to ductile, power law distribution of the damage events, shear bands strain localization, fractal structure of the damage. These properties are not incorporated at the elementary scale but results from the interaction between elements. The emergence of these macroscopic properties allows us to consider the rock deformation process as a complex system displaying non-linear dynamics.

## 1. Introduction

Le comportement mécanique des roches, pour des échelles allant de celle de l'échantillon de laboratoire à celle de la croûte terrestre, est souvent décrit comme complexe, terme couramment employé comme synonyme de compliqué. Cette complexité tient d'une part à la diversité des comportements mécaniques observés pour les roches, (élastique/plastique, fragile/ductile, endommageant, visqueux,…) et d'autre part aux nombreux paramètres influençant le comportement mécanique. Il peut s'agir de paramètres liés à la composition du matériau rocheux (minéralogie, microstructure, densité, distribution de fissuration), ou aux conditions de chargement (pression, température, vitesse de déformation). Ainsi un même matériau pourra montrer un comportement fragile ou ductile selon les conditions de chargement [13]. La complexité tient également aux processus de déformations à petite échelle (propagation de fissures, plasticité par saut de dislocation, par glissement sur des discontinuités) qui interagissent entre eux et dont il est difficile d'estimer l'effet sur le comportement mécanique macroscopique.

La complexité du comportement mécanique apparaît également au sein même du processus de déformation. Lors de la sollicitation mécanique du matériau rocheux, la déformation globale comporte des déformations inélastiques localisées qui peuvent être quantifiées par le biais de l'onde élastique de déformation et de contrainte émise [18, 21]. On l'observe aussi bien à l'échelle crustale (sismicité), qu'à l'échelle des massifs rocheux (microsismicité) qu'à celle des échantillons de laboratoire (émission acoustique, EA). A ces différentes échelles, la sismicité montre à la fois une imprédictibilité de chaque évènement pris individuellement (taille, lieu et temps d'occurrence) et une prédictibilité statistique (raccordement à des lois statistiques bien identifiées). En effet, on observe des distributions en loi puissance particulièrement robustes dans les trois domaines de la taille des événements sismiques [8, 18-22], de leur répartition spatiale [12, 17, 22, 25, 26] et temporelle [11, 17, 23]. Ces distributions indiquent une invariance d'échelle c'est à dire une absence de taille caractéristique [20] et une corrélation entre différentes échelles.

Des expériences de laboratoire ont permis d'observer des variations de ces paramètres corrélées aux différentes phases du comportement mécanique macroscopique [1, 12 ,18 ,21]. Par ailleurs, il a été observé que l'invariance d'échelle peut être limitée soit par la taille finie de l'objet considéré [19, 25], soit par une taille caractéristique de sa structure interne [25], soit encore par un changement des mécanismes de rupture selon l'échelle considérée [19].

Enfin, l'observation de l'état d'endommagement de roches après déformation mécanique révèle des propriétés fractales (rugosité des surfaces de rupture, microstructure, …) qui sont des indicateurs supplémentaires de la complexité [3, 9, 15, 24].

Pour mieux comprendre cette complexité, une voie consiste à intégrer l'ensemble des paramètres jugés pertinents dans le comportement élémentaire (propriétés mécaniques, processus micro-mécanique, micro-structure) et à simuler le comportement mécanique ainsi obtenu. Cette approche se confronte à de nombreuses difficultés à la fois méthodologiques (estimations des paramètres, hiérarchisations des processus, description effective de la structure) et numériques (absence de méthode générale pour prendre en compte l'ensemble des processus).

Une autre voie consiste à utiliser une loi élémentaire simple et à faire interagir un grand nombre d'éléments. Des modèles numériques de ce type ont été proposés qui permettent de simuler un comportement macroscopique varié (du ductile au fragile) à partir d'un comportement élémentaire simple [1, 2, 27]. Nous proposons ici de recourir à ce type de modèle pour montrer qu'un certain nombre d'observations relatives au comportement mécanique des roches, peuvent être vues comme des propriétés émergentes d'un système complexe. Le terme complexe est utilisé ici comme définissant un système dont les propriétés macroscopiques sont absentes à l'échelle élémentaire et émergent de l'interaction de ses éléments [4, 16].

## 2. Description du modèle numérique

Le modèle utilisé est basé sur un endommagement élastique progressif isotrope. Le module élastique effectif du matériau endommagé, $E_{eff}$, est fonction du module élastique intial, $E_0$, et du paramètre d'endommagement, D.

$$E_{eff.} = E_0.(1-D) \qquad (1)$$

D peut être quantitativement relié à la densité de fissure [14]. Cette relation suppose que la taille d'un élément est grande devant celle des défauts qu'il contient. L'échelle élémentaire est donc mésoscopique. La résolution de l'état mécanique (contrainte déformation) se fait par la méthode des éléments finis en 2D sous l'hypothèse des déformations planes. Le modèle macroscopique est chargé en augmentant progressivement le déplacement imposé à la limite supérieure du modèle. Lorsqu'un élément atteint son seuil d'endommagement, son module élastique est diminué par un facteur (1-D). Du fait des redistributions de contrainte induites par cet endommagement sur les éléments environnants, une avalanche peut se déclencher. Le nombre d'endommagement au cours d'un pas de chargement constitue la taille de l'avalanche.

Le seuil d'endommagement est déterminé par le critère de Mohr-Coulomb :

$$\tau = \sigma \tan \phi + C \qquad (2)$$

Ou $\tau$ est la contrainte tangentielle ; $\sigma$ est la contrainte normale ; C est la cohésion et $\phi$ est l'angle de frottement interne. Pour un modèle entièrement homogène, le comportement élémentaire est reproduit à l'identique à l'échelle macroscopique [1, 2]. Au contraire, la présence d'hétérogénéités conduit à l'émergence d'un comportement macroscopique différent du comportement élémentaire. L'hétérogénéité du matériau est simulée en attribuant la cohésion aléatoirement selon une distribution homogène. Pour plus de détails sur le modèle numérique, voir les références [1, 2]

## 3. Propriétés macroscopiques des simulations

L'étude de sensibilité du modèle aux différents paramètres a montré que le comportement macroscopique est uniquement déterminé par le paramètre $\phi$ [1, 2]. La figure 1 présente les résultats de simulations pour $\phi$ variant de 0 à 56°. On présente le comportement mécanique, l'état d'endommagement final, la distribution de la taille des avalanches et l'intégrale de corrélation spatiale de l'endommagement, C(r) [6]. La dimension de corrélation spatiale $D_2$ est déterminée en estimant la pente de C(r) dans une représentation loglog. L'ajustement est réalisé uniquement dans la section en loi puissance de C(r).

A l'échelle macroscopique (modèle pris dans son ensemble), le comportement mécanique comprend une phase linéaire, puis l'apparition d'une non-linéarité associée à l'apparition d'avalanches. Selon la valeur du paramètre $\phi$, le comportement peut varier continûment du ductile au fragile (figure 1a). Dans le cas fragile, la contrainte macroscopique connaît une chute importante et brutale associée à une localisation soudaine de l'endommagement. Dans le cas ductile, la contrainte atteint une valeur stable sans chute de contrainte. L'endommagement se localise progressivement sous forme de bande (figure 1b). Toutes les situations intermédiaires entre ces deux cas extrêmes peuvent être simulées en faisant varier le paramètre $\phi$. La distribution de la taille des événements suit une loi puissance dont l'exposant dépend de $\phi$ (figure 1c). L'intégrale de corrélation spatiale C (r) suit également une loi puissance (figure 1d), ce qui indique que l'endommagement possède une structure fractale. La dimension de corrélation spatiale $D_2$ (pente de l'intégrale en diagramme loglog) dépend également de $\phi$.

L'ensemble des propriétés macroscopiques est sensible au paramètre $\phi$. L'étude du champ de contrainte autour de défauts a montré que ce paramètre conditionne fortement la géométrie d'interaction entre les éléments [1, 2]. Celle-ci est d'autant plus anisotrope que $\phi$ est élevé. Cette anisotropie influe à la fois sur la localisation de l'endommagement, sur la dynamique des avalanches et sur le comportement macroscopique. Ces résultats sont obtenus avec un modèle à deux dimensions. Etant donné le rôle prépondérant de la géométrie d'interaction sur les résultats de simulation, on peut considérer que le passage à trois dimensions pourrait avoir un effet important. Pour des modèles d'endommagement de réseau de fibre [9], proches du modèle présenté ici, le passage de 2D à 3D n'induit pas de changement significatif sur le comportement macroscopique ni sur l'émergence de distribution en loi puissance. Par contre, cela change significativement les valeurs des exposants caractérisant ces lois puissance.

**4. Emergence de la complexité et implications pour la dynamique de la déformation des roches**

Le modèle proposé est basé sur un endommagement élémentaire progressif dans un modèle élastique hétérogène. Le comportement élémentaire est très simple : diminution progressive du module élastique sans perte de résistance. Chaque élément possède des propriétés isotropes et ne connaît que des événements d'endommagement de taille unitaire. A l'échelle macroscopique, on observe un comportement complexe à plusieurs titres : le comportement mécanique est non linéaire et peut varier du ductile au fragile ; l'endommagement final possède une structure fractale plus ou moins localisée ; les avalanches suivent une distribution en loi de puissance. La figure 2 synthétise l'ensemble de ces observations. Ces propriétés n'étant pas incorporées explicitement dans le comportement élémentaire, il s'agit de propriétés émergentes du système résultant de l'interaction entre éléments. Nous avons en outre montré que la modification de la géométrie d'interaction influence l'ensemble des propriétés macroscopiques. En cela, le processus de déformation s'apparente à un système complexe [4, 10, 16]. L'émergence d'invariance d'échelle dans la taille et la répartition spatiale est un aspect supplémentaire de la complexité [4].

Le type de modèle présenté ici permet de rapprocher l'étude du comportement mécanique des roches de celle d'autres systèmes dynamiques (tas de sable, percolation, ising, blocs-ressorts,… ) possédant des propriétés particulières (critique, critique auto organisé,…) qui peuvent être étudiées par les outils de la physique statistique [5, 7, 16, 27]. En cela, cette approche constitue une alternative à l'approche mécanique classique (approche linéaire) qui considère le comportement mécanique macroscopique comme la somme des comportements élémentaires.

## 5. Conclusion

Nous avons présenté un modèle mécanique simple pour le comportement des roches. Celui-ci permet de simuler plusieurs observations expérimentales : comportement mécanique allant du fragile au ductile, distribution en loi puissance des événements d'endommagement, localisation de la déformation sous forme de bande, structure fractale de l'endommagement.

Ces propriétés ne sont pas introduites à l'échelle élémentaire et résultent de l'interaction entre les éléments du modèle. L'émergence de ces propriétés macroscopiques permet de considérer le processus de déformation des roches comme un système complexe possédant une dynamique non-linéaire.

## 6. Références

## 7. Remerciements

**Figures :**

**Figure 1** : Résultats de simulation pour différentes valeurs de $\phi$. a) Comportement mécanique macroscopique, b) endommagement final du modèle, c) distribution de la taille des avalanches, d) intégrale de corrélation spatiale de l'endommagement.

Figure 1 : Simulation results for various values of $\phi$. a) Macroscopic mechanical behavior b) Final damage state, c) Avalanches size-frequency, d) Spatial correlation integral of the damage.

**Figure 2** : Emergence de la complexité. a) Echelle élémentaire et macroscopique, b) Comportement mécanique élémentaire et macroscopique, c) Localisation de l'endommagement, d) Distribution de la taille des avalanches.

Figure 2 : Emerging complexity. a) Elementary and macroscopic scales, b) Elementary and macroscopic mechanical behavior, c) Damage localization, d) Avalanches size frequency.

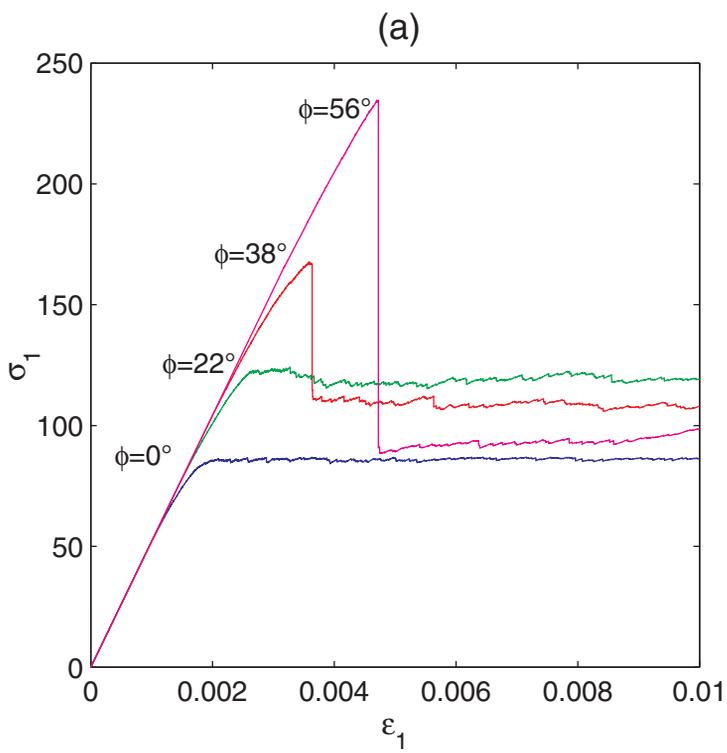
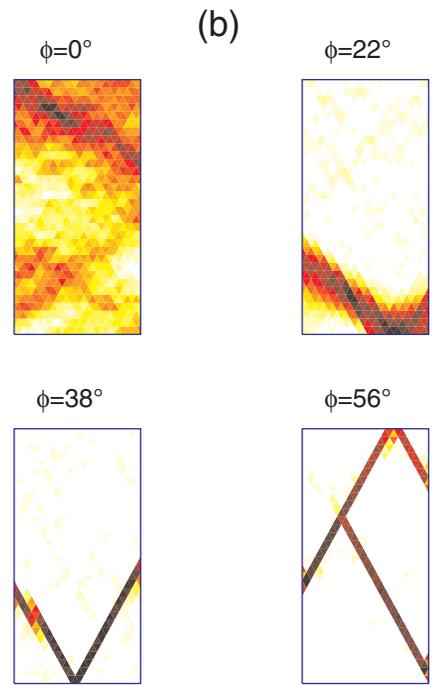
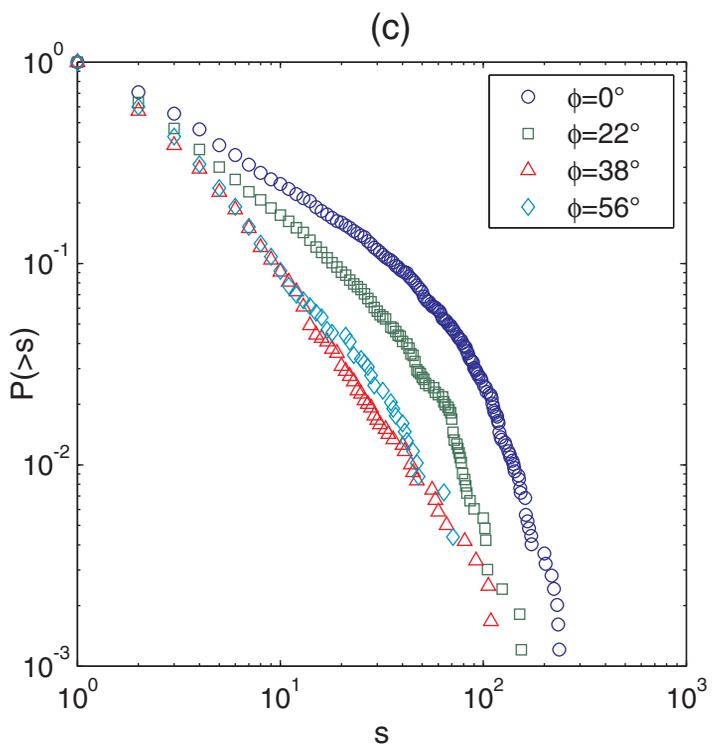
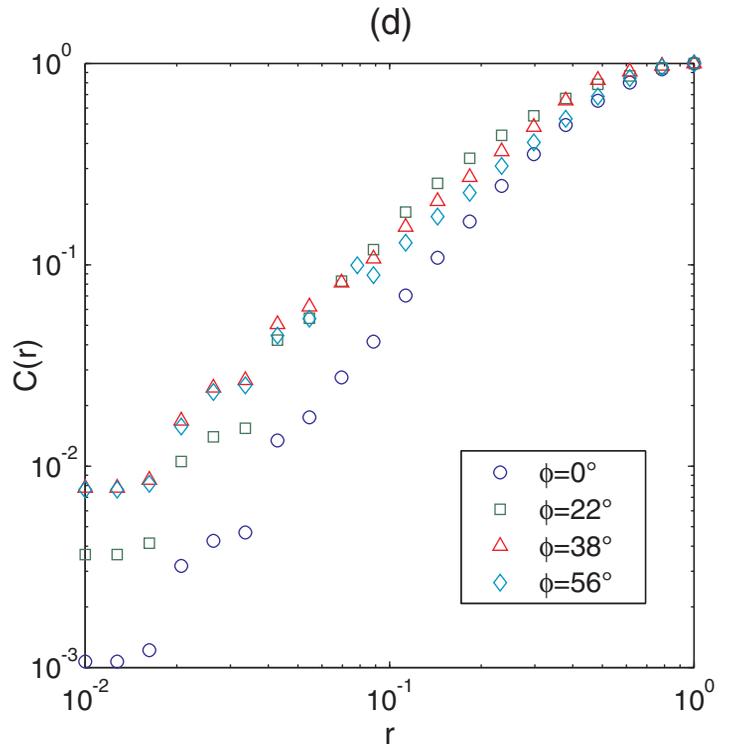

Figure 1

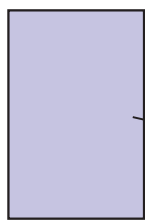
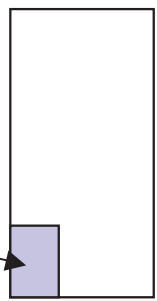

(a) Élément homogène — Échelle mesoscopique — Interaction élastique + hétérogénéité — Modèle Échelle macroscopique

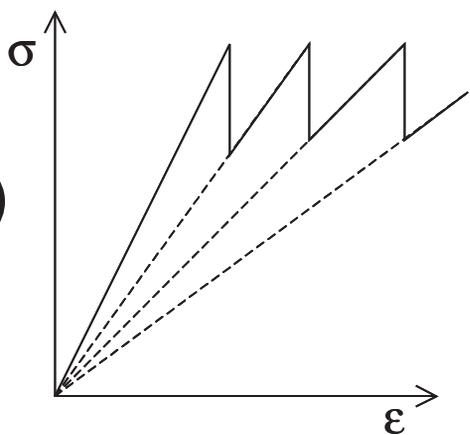
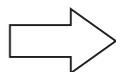
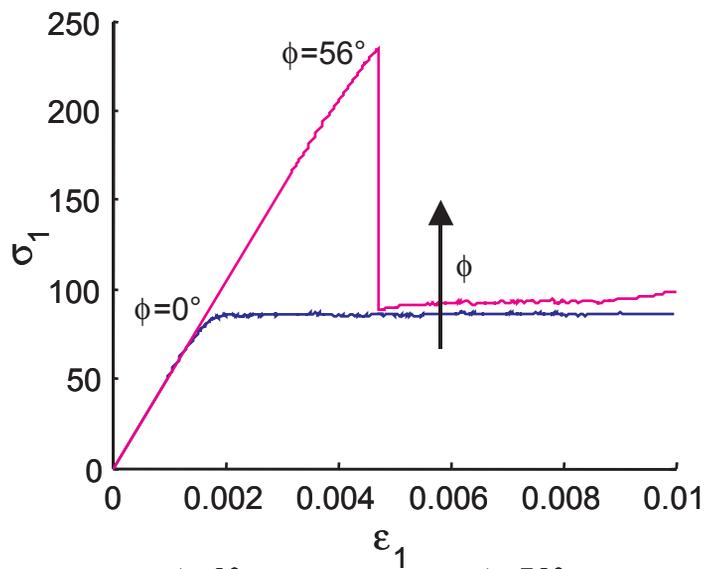

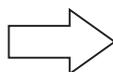
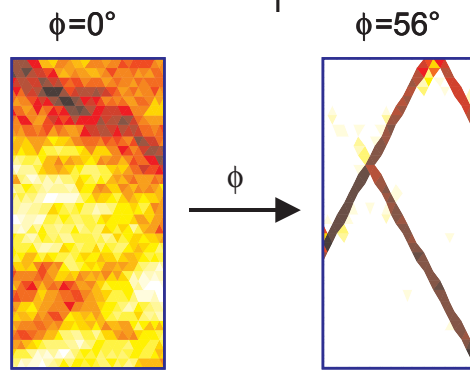

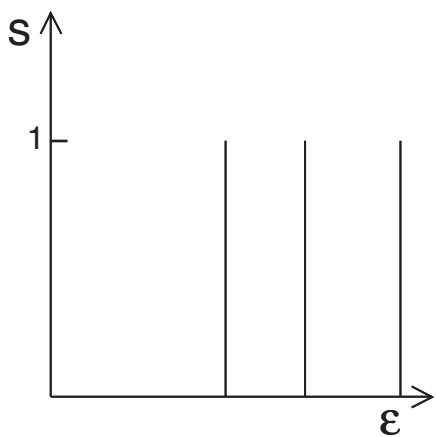
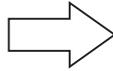
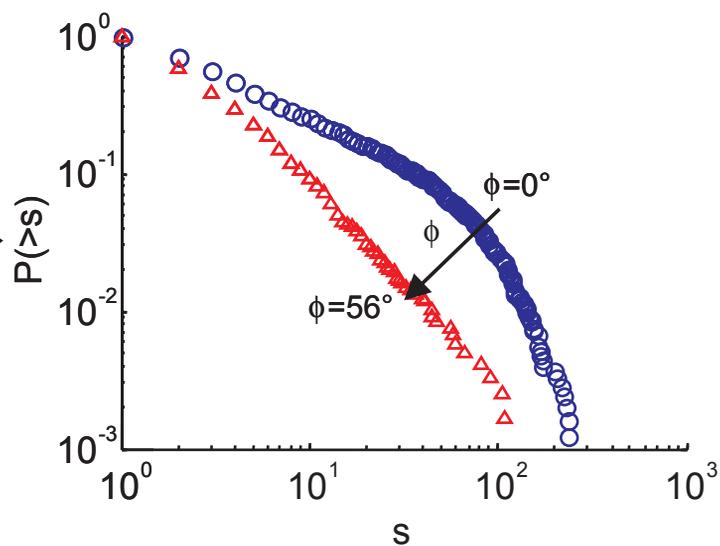